\documentclass[11pt]{article}

\pdfoutput=1

\usepackage[T1]{fontenc}
\usepackage[latin9]{inputenc}
\usepackage[a4paper]{geometry}
\usepackage[active]{srcltx}
\usepackage{amsmath}
\usepackage{amssymb}
\usepackage{esint}
\usepackage{ulem}

\usepackage{xcolor}

\makeatletter
\usepackage{textcomp}

\pdfoutput=1 % if your are submitting a pdflatex (i.e. if you have
\usepackage{jheppub}% for details on the use of the package, please
\newcommand*\xbar[1]{%
  \hbox{%
    \vbox{%
      \hrule height 0.5pt % The actual bar
      \kern0.3ex%         % Distance between bar and symbol
      \hbox{%
        \kern-0.0em%      % Shortening on the left side
        \ensuremath{#1}%
        \kern-0.0em%      % Shortening on the right side
      }%
    }%
  }%
}

\usepackage{amsfonts}

\usepackage{bbm}

\newcommand{\be}{\begin{equation}}
\newcommand{\ee}{\end{equation}}
\newcommand{\bea}{\begin{eqnarray}}
\newcommand{\eea}{\end{eqnarray}}

\title{\boldmath Simplifying (super-)BMS algebras}

\author[a]{Oscar Fuentealba,}
\author[a,b]{Marc Henneaux}

\affiliation[a]{Universit\'e Libre de Bruxelles and International Solvay Institutes, ULB-Campus Plaine CP231, B-1050 Brussels, Belgium}
\affiliation[b]{Coll\`ege de France,  Universit\'e PSL, 11 place Marcelin Berthelot, 75005 Paris, France}

\emailAdd{oscar.fuentealba@ulb.be}
\emailAdd{marc.henneaux@ulb.be}

\preprint{}

\abstract{We show that the non-linear BMS$_5$ symmetry algebra of asymptotically flat Einstein gravity in five dimensions, as well as the super-BMS$_4$ superalgebra of asymptotically flat supergravity, can be redefined so as to take a direct sum structure.  In the new presentation of the (super-)algebra, angle-dependent translations and angle-dependent supersymmetry transformations commute with the (super-)Poincar\'e generators.  We also explain in detail the structure and charge-integrability of asymptotic symmetries with symmetry parameters depending on the fields (through the charges themselves), a topic relevant for nonlinear asymptotic symmetry algebras.}

\makeatother

\begin{document}
\maketitle \flushbottom

\section{Introduction}

In a recent paper \cite{Fuentealba:2022xsz}, we showed how pure BMS$_4$ supertranslations \cite{Bondi:1962px,Sachs:1962wk,Sachs:1962zza}  could be made to commute with all Poincar\'e transformations by introducing logarithmic supertranslations and allowing non-linear redefinitions of the homogeneous Lorentz charge-generators. In retrospect, this is perhaps not too surprizing since Poincar\'e transformations and pure supertranslations have a different origin when one traces them back to the linear theory \cite{Fuentealba:2020ghw}: while the Poincar\'e transformations are rigid symmetries of the linear theory, with a non-trivial bulk contribution to their charge-generator, the pure supertranslations appear already there as ``improper \cite{Benguria:1976in} gauge transformations''.  This decoupling of the symmetries in the algebra, which becomes a direct sum,  is also in the spirit of the Coleman-Mandula theorem \cite{Coleman:1967ad}.

The redefinitions of the generators of the BMS$_4$ algebra shed furthermore new light on what is sometimes called the ``angular momentum ambiguity'' since they provide a supertranslation independent definition of the Poincar\'e charges.  It was proved in \cite{Fuentealba:2023syb} that the redefinitions of \cite{Fuentealba:2022xsz} made at spatial infinity and those of 
\cite{Mirbabayi:2016axw,Bousso:2017dny,Javadinezhad:2018urv,Javadinezhad:2022hhl,Chen:2021szm,Chen:2021zmu,Compere:2021inq,Compere:2023qoa} carried out at null infinity were equivalent.

What makes the algebraic decoupling possible is that the generators of the logarithmic supertranslations are canonically conjugate to those of the pure supertranslations, allowing to eliminate by a Darboux-like procedure cross-terms in the brackets between the generators of the pure supertranslations with the other generators of the BMS$_4$ algebra.  The charges generating logarithmic supertranslations were shown in \cite{Fuentealba:2023syb} to match the null infinity potential $C$ for the electric part of the Bondi shear, identified in \cite{Strominger:2013jfa,Strominger:2017zoo} to be the Goldstone boson of spontaneously broken supertranslation invariance.  The importance of the logarithmic branch to describe complete canonical pairs including the Goldstone boson was  derived in a different context, at null infinity,  in  \cite{Donnay:2018neh}.

The purpose of this article is to extend the construction of  \cite{Fuentealba:2022xsz} to BMS algebras in higher dimensions -- specifically in the explicit case of 5 dimensions \cite{Fuentealba:2021yvo,Fuentealba:2022yqt} -- as well as to the BMS$_4$ superalgebra of \cite{Fuentealba:2021xhn}.  What makes the extension and the algebraic decoupling possible is that in these algebras, the generators of pure supertranslations or pure angle-dependent supersymmetry transformations have also canonical partners, just as in the case of the log-BMS$_4$ algebra.  These conjugate variables can be used to perform the appropriate redefinitions of the (super-)Poincar\'e generators, again along Darboux-like lines. These redefinitions are nonlinear.  For that reason, we also explain in detail the structure of the corresponding nonlinear charges and the asymptotic symmetry transformations that they generate. 

Our paper is organized as follows. In Section {\bf \ref{Sec:AsymptoticNonlinear}}, we {review the definition of asymptotic symmetries and recall  how nonlinear redefinitions of the charges fit into the construction in a manner that preserves the general local structure (``weakly vanishing bulk term $+$ surface integral at infinity'') of the generators of asymptotic symmetries.}   We  consider next  in Section {\bf \ref{sec:BMS5}} the BMS$_5$ algebra and show that one can algebraically decouple the Poincar\'e subalgebra from the pure supertranslations and their subleading nontrivial partners. In Section {\bf \ref{Sec:SUSYV1}}, we turn to supersymmetry and consider first the superalgebra of \cite{Fuentealba:2021xhn}, where we show how to decouple the super-BMS generators in the superalgebra.  By introducing the logarithmic supertranslations in the superalgebra of \cite{Fuentealba:2021xhn}, we {define the log-BMS$_4$ superalgebra in Section {\bf \ref{Sec:SUSYV2}} and achieve}  a complete algebraic decoupling.  We show that the log-BMS$_4$ superalgebra can indeed be redefined so as to take the direct sum structure of the finite-dimensional super-Poincar\'e algebra with the infinite-dimensional superalgebra containing the pure BMS supertranslations, the logarithmic supertranslations and their fermionic partners.  We close our paper with some comments on generalizations (Section {\bf \ref{sec:Conclusions}}) and a technical appendix.

Finally, a note on notations:  in the pure bosonic case, Poisson brackets will be denoted by $\{, \}$.  When we deal with supersymmetric theories, however, we shall use a mixed notation ($[,]$ for the Poisson bracket between a bosonic variable and a bosonic or fermionic one,  $\{, \}$ for the Poisson bracket between two fermionic variables) to emphasize the symmetry properties.

 \section{Asymptotic symmetries and nonlinear redefinitions}
 \label{Sec:AsymptoticNonlinear}

\subsection{Asymptotic symmetries}
Asymptotic symmetries are gauge transformations that preserve the boundary conditions and the action.  They are thus canonical transformations with well-defined generators, which we call equivalently ``charge-generators'', ``generators'' or ``charges''.  Denoting the (first class) constraints by $\mathcal H_\alpha \approx 0$ and the gauge parameters by $\xi^\alpha$, the generators of asymptotic symmetries take the form
\be
G[\xi^\alpha] = \int d^d x \xi^\alpha \mathcal H_\alpha +B_\xi \,, \label{Eq:Generators0}
\ee
where $B_\xi = \oint_{S^{d-1}_\infty} d^{d-1}y f$ is an appropriate surface term evaluated on the $(d-1)$-sphere $S^{d-1}_\infty$ at infinity, which is necessary to make the key formula $\iota_{X_\xi} \Omega = - d_V G[\xi^\alpha]$ hold. The phace space vector field $X_\xi$ is the Hamiltonian vector field associated with the generator $G[\xi^\alpha]$, the phase space $2$-form $\Omega$ is the symplectic form and $d_V$ is the phase space exterior derivative, not to be confused with the exterior derivative in space(time) $d$ (see \cite{Henneaux:2018gfi} for more information). The coordinates $y^A$ are coordinates on the $(d-1)$-sphere, and polar coordinates are $(x^k) =(r, y^A)$. The integrand $f$ of the surface integral giving $B_\xi$ depends on the asymptotic values of the gauge parameters and of the fields.  

If the gauge parameters decrease fast enough at infinity, in a way that makes the boundary term $B_\xi$ equal to zero, the generator $G[\xi^\alpha]$ weakly vanishes,  $G[\xi^\alpha]\approx 0$, and the corresponding transformation is a proper gauge transformation implying a redundancy in the description of the system \cite{Benguria:1976in}.  If the boundary term $B_\xi$ does not vanish,  $G[\xi^\alpha]$ generates an improper gauge transformation with physical content which should not be factored.  Two generators $G[\xi^\alpha]$ and $G[\xi'^\alpha]$ that differ by constraint terms, i.e., which are weakly equal, 
\be
G[\xi^\alpha] \approx G[\xi'^\alpha]  \label{Eq:Equivalence}
\ee
generate transformations that coincide at infinity and differ only in the bulk, by proper gauge transformations.  They are thus physically equivalent. The physical content of asymptotic symmetries is entirely captured by their asymptotic behaviour, hence the terminology.  Note that the generators $G[\xi^\alpha]$ are invariant under proper gauge transformations, even if the asymptotic symmetry algebra is non-abelian.

We will carry on the redundancy (\ref{Eq:Equivalence}) and will thus work below ``up to proper gauge transformations'', i.e., up to weakly vanishing terms for the gauge-invariant generators (even if not explicitly stated).  For that reason, we shall often only write down explicitly the asymptotic expression of the charges, i.e., focus only on the relevant boundary terms.  Equivalently, one could fix the proper gauge by proper gauge fixing conditions and introduce the Dirac bracket.  As it is known, the Dirac brackets of (proper) gauge-invariant quantities coincide with their Poisson brackets (see e.g. \cite{Henneaux:1992ig}), so the two methods are indeed equivalent.

The asymptotic symmetries depend on the asymptotic values of the gauge parameters at infinity, 
which are parametrized in the general case by constant parameters $U^s$ (e.g., Poincar\'e transformations) and functions on the sphere at infinity, which we denote by $T^a(y)$ (e.g., supertranslations).  One has, for the gauge parameters $\xi^\alpha$, the asymptotic expansion
\be
\xi^\alpha (r,y)\rightarrow_{r \rightarrow \infty} \overset{\circ}{\xi}\,^\alpha (r, y^A, U^s, T^a, \partial_A T ^a, \cdots) + \textrm{``more''}  \,,\label{Eq:GaugePara2}
\ee
where $\overset{\circ}{\xi}\,^\alpha$ is the leading term which takes a specific form (at leading order) in terms of $r$, $y^A$, of the asymptotic values of the fields as well as of the asymptotic symmetry parameters $U^s$ and $T^a(y)$ and their derivatives up to some order.  The parameters and their derivatives  occur linearly.   The term  ``more'' in (\ref{Eq:GaugePara2}) denotes subleading terms irrelevant for the surface integrals at infinity.    The charge-generators of the gauge transformations then take the form 
\be
G[\xi^\alpha] = \int d^d x \xi^\alpha \mathcal H_\alpha + U^s \oint d^{d-1}y  \mathcal Q_s + \oint d^{d-1}y T^a \mathcal G_a \,. \label{Eq:Generators1}
\ee
We assume that the parametrization of the asymptotic symmetries is chosen such that $U^s$ and $T^a(y)$ do not depend on the fields and that the charge-generators $G[\xi^\alpha] $ given by (\ref{Eq:Generators1}) have therefore well-defined functional derivatives under that condition.   We will investigate below what happens when one makes redefinitions involving the fields (through the charges). 

The asymptotic charges are, modulo unwritten weakly vanishing bulk terms,
\be
Q_s = \oint d^{d-1}y  \mathcal Q_s\,, \qquad Q_a (y) =  \mathcal G_a(y)\,.
\ee
The Poisson brackets $\{Q_s, Q_r\}$, $\{Q_s, Q_a(y)\}$ and $\{Q_a(y), Q_b(y')\}$ of the charges 
are on general grounds functions of the charges, which might be nonlinear.  Nonlinear algebras are in fact the rule rather than a fancy exception (see e.g.  \cite{deBoer:1995cqx}).  Explicit examples in the context of asymptotic symmetries and charges have been derived in \cite{Henneaux:1999ib,Henneaux:2010xg,Campoleoni:2010zq,Afshar:2013vka,Gonzalez:2013oaa,Henneaux:2015ywa,Fuentealba:2015wza,Henneaux:2015tar,Fuentealba:2020zkf}.

It must be stressed that the nonlinear functions of the charges occuring in the brackets or in the redefinitions made below are still given by expressions such as (\ref{Eq:Generators0}) with appropriate gauge parameters and boundary terms and not, say, by nonlocal expressions such as $( \int d^d x \xi^\alpha \mathcal H_\alpha +B_\xi)^2$.  Since there might be some confusion on that point, we shall develop it further.

\subsection{Equations obeyed by the charges}

We denote the canonically conjugate fields by $(\phi^\Gamma, \pi_\Gamma)$ and  assume for definiteness and simplicity that  the symplectic structure is
\be
\Omega = \int d^d x \, d_V \pi_\Gamma \wedge d_V \phi^\Gamma.
\ee
In that case, the condition $\iota_{X_M} \Omega = -d_V M$ for $M = \int d^dx \mathcal M + \oint d^{d-1} y \, m$  implies the condition of  \cite{Regge:1974zd} that $M$ should have well defined functional derivatives, i.e., that 
$$ d_V M = \int d^d x \Big(L_\Gamma d_V \phi^\Gamma + N^\Gamma d_V \pi_\Gamma\Big)\,, $$
for some $L_\Gamma$ and $N^\Gamma$ (without extra surface terms since there is no differentiated $d_V \pi_\Gamma$ or  $d_V \phi^\Gamma$ in $\Omega$), which are identified as the functional derivatives of $M$,
$$
\frac{\delta M}{\delta \phi^\Gamma(x)} = L_\Gamma(x)\,, \quad \frac{\delta M}{\delta \pi_\Gamma(x)} = N^\Gamma(x)\,.$$
  The corresponding phase space Hamiltonian vector field has components
$$
X_M = (N^\Gamma, - L_\Gamma) \quad \Leftrightarrow \quad X_M = \int d^dx \left( \frac{\delta M}{\delta \pi_\Gamma(x)}\frac{\delta }{\delta \phi^\Gamma(x)}- \frac{\delta M}{\delta \phi^\Gamma(x)}\frac{\delta }{\delta \pi_\Gamma(x)}\right)\,.
$$
With a more general symplectic structure having for instance additional surface contributions, the condition $\iota_{X_M} \Omega = -d_V M$ still holds but implies conditions that generalize those of \cite{Regge:1974zd}, as explained in  \cite{Henneaux:2018gfi}.

Let us compute $d_V G[\xi^\alpha]$, with $d_V U^s= d_V T^a(y) = 0$.  We get
\be
d_V G[\xi^\alpha] = \int d^d x (d_V\xi^\alpha) \mathcal H_\alpha + \int d^d x \xi^\alpha d_V \mathcal H_\alpha+ U^s \oint d^{d-1}y  d_V\mathcal Q_s + \oint d^{d-1}y T^a d_V \mathcal G_a\,,
\ee
where we allow for a field-dependence of $\xi^\alpha$ in the bulk (compatible with its asymptotic behaviour).  Now, 
$\int d^d x \xi^\alpha d_V \mathcal H_\alpha$ can be transformed by  integration by parts into the bulk form $\int d^d x A_\xi^\Gamma d_V \pi_\Gamma - \int d^d x d_V \phi^\Gamma B_{\xi,\Gamma}$ with undifferentiated $d_V \phi^\Gamma$'s and $d_V \pi_\Gamma$'s, plus surface terms.  Keeping all terms - i.e., making no use of the boundary conditions at this stage -,  yields an identity of the form
\be
\int d^d x \xi^\alpha d_V \mathcal H_\alpha = \int d^d x A_\xi^\Gamma d_V \pi_\Gamma - \int d^d x d_V \phi^\Gamma B_{\xi,\Gamma} + \oint d^{d-1}y \mathcal V \,,\label{Iden1}
\ee
for some $A_\xi^\Gamma$ and $B_{\xi,\Gamma}$.  Here, the integrand $\mathcal V$ of the surface integral at infinity  involves the asymptotic value  of $\xi_\alpha$, as well as $d_V \phi^\Gamma$,  $d_V \pi_\Gamma$  and possibly their derivatives.   
The gauge parameters $U^s$, $T^a(y)$ occur linearly through $\overset{\circ}{\xi}\,^\alpha$. By integrating by parts on the sphere, one can remove the derivatives of $T_a$ (if any), so that $\oint d^{d-1}y \mathcal V$ is identical to,
\be
\oint d^{d-1}y \mathcal V = U^s \oint d^{d-1}y k_s + \oint d^{d-1}y T^a s_a\,.  \label{Iden2}
\ee
Similarly, one gets another integration-by-part identity, 
\be
\int d^d x (d_V\xi^\alpha) \mathcal H_\alpha = \int d^d x {A'}_\xi^\Gamma d_V \pi_\Gamma - \int d^d x d_V \phi^\Gamma B'_{\xi,\Gamma} +  \oint d^{d-1}y \mathcal V' \,,  \label{Iden3}
\ee
for some ${A'}_\xi^\Gamma$, $B'_{\xi,\Gamma}$ and $\mathcal V'$, which weakly vanish since they are proportional to the constraints.  One has again
\be
\oint d^{d-1}y \mathcal V' = U^s \oint d^{d-1}y k'_s + \oint d^{d-1}y T^a s'_a  \,.  \label{Iden4}
\ee

The fact that $G[\xi^\alpha]$ has well-defined functional derivatives, i.e., $d_V G[\xi^\alpha]$ reduces to the bulk terms, is equivalent to the condition
\be
\oint d^{d-1}y \mathcal V' + \oint d^{d-1}y \mathcal V + U^s \oint d^{d-1}y  d_V\mathcal Q_s + \oint d^{d-1}y T^a d_V \mathcal G_a = 0\,. \label{Eq:0ST}
\ee
While the relations (\ref{Iden1})-(\ref{Iden4}) are identities, the equation (\ref{Eq:0ST}) is in general not identically satisfied but holds only if use is made of the boundary conditions.  
One can rewrite it as
\be
U^s \oint d^{d-1}y  (d_V \mathcal Q_s + k_s + k'_s)+ \oint d^{d-1}y T^a (d_V \mathcal G_a + s_a + s'_a) = 0\,.
\ee
Since the $U^s$'s and the $T^a$'s are arbitrary, this is equivalent to\footnote{If the $T^a(y)$'s are not arbitrary functions on the sphere but are restricted, say, by parity conditions, the conditions (\ref{Eq:EqForQG}) should be understood as ``modulo terms that project to zero when enforcing the relevant functional space restrictions''.}
\be
d_V Q_s + \oint d^{d-1}y  (k_s + k'_s) = 0\,, \qquad d_V \mathcal G_a + s_a + s'_a = 0\,.  \label{Eq:EqForQG}
\ee
Note that in many cases, the boundary conditions enforce that the constraints decrease sufficiently fast at infinity that they do not contribute to the relevant surface integrals, so that $k'_s= s'_a(y) = 0$, but this is not necessary for our discussion.

\subsection{Nonlinear charges}

The question we examine in this subsection is how to determine the complete (bulk + surface terms) form of the generators of asymptotic symmetries when one prescribes the surface terms to be some specific function of the charges.  That is, what is the $\xbar \xi^\alpha$ to be included in
$
G[\xbar \xi^\alpha] = \int d^d x \xbar \xi^\alpha \mathcal H_\alpha + F[Q_s, Q_a(y)]
$
where $F[Q_s, Q_a(y)]$ is some given functional of the charges?

We define 
\be
\xbar U^s = \frac {\partial F}{\partial Q_s}  \,, \qquad 
\xbar T^a(y) = \frac {\delta F}{\delta Q_a(y)} \,,
\ee
and consider now gauge parameters $\xbar \xi^\alpha$ that go asymptotically to 
\be
\xbar \xi^\alpha \quad \rightarrow_{r \rightarrow \infty} \quad \overset{\circ}{\xi}\,^\alpha (r, y^A, \xbar U^s, \xbar T^a, \partial_A \xbar T ^a, \cdots) 
\ee in which the asymptotic gauge parameters $U^s$, $T^a$ are replaced by $\xbar U^s$, $\xbar T^a$.   Such gauge parameters $\xbar \xi^\alpha$ are not unique but the ambiguity corresponds to a proper gauge transformation and is physically irrelevant. We claim that the expression
\be
G[\xbar \xi^\alpha] = \int d^d x \xbar \xi^\alpha \mathcal H_\alpha + F[Q_s, Q_a(y)] \, ,
\ee
with that choice of $\xbar \xi^\alpha$, is a well-defined generator, with charge clearly (weakly) equal to $ F$. 

To establish this fact, we note that one has 
\be
 d_V G[\xbar \xi^\alpha] = \int d^d x (d_V \xbar \xi^\alpha) \mathcal H_\alpha + \int d^d x \xbar \xi^\alpha d_V \mathcal H_\alpha + d_V F[Q_s, Q_a(y)] \,. 
 \ee
 Using the identities (\ref{Iden1})-(\ref{Iden4}), one can bring the first two terms of the right-hand side to the form
 \begin{align}
  \int d^d x (d_V \xbar \xi^\alpha) \mathcal H_\alpha + \int d^d x \xbar \xi^\alpha d_V \mathcal H_\alpha & = 
 \int d^d x \left((A_{\xbar \xi}^\Gamma + {A'}_{\xbar \xi}^\Gamma) d_V \pi_\Gamma -  d_V \phi^\Gamma (B_{\xbar \xi,\Gamma} + B'_{\xbar \xi,\Gamma} ) \right) \nonumber \\
 & + \xbar U^s \oint d^{d-1}y (k_s + k'_s)+ \oint d^{d-1}y \xbar T^a (s_a + s'_a)\,,
\end{align}
which, using the equality (\ref{Eq:EqForQG}), can be rewritten as
\begin{align}
  \int d^d x (d_V \xbar \xi^\alpha) \mathcal H_\alpha + \int d^d x \xbar \xi^\alpha d_V \mathcal H_\alpha & = 
 \int d^d x \left((A_{\xbar \xi}^\Gamma + {A'}_{\xbar \xi}^\Gamma) d_V \pi_\Gamma -  d_V \phi^\Gamma (B_{\xbar \xi,\Gamma} + B'_{\xbar \xi,\Gamma} ) \right) \nonumber \\
 & - \xbar U^s d_V Q_s - \oint d^{d-1}y \xbar T^a d_V \mathcal G_a\,.
\end{align}
On the other hand, $d_V F[Q_s, Q_a(y)]$ is equal to 
\be
d_V F[Q_s, Q_a(y)] = \frac {\partial F}{\partial Q_s} d_V Q_s + \oint d^{d-1}y  \frac {\delta F}{\delta Q_a(y)} d_V \mathcal G_a = \xbar U^s d_V Q_s + \oint d^{d-1}y  \xbar T^a d_V \mathcal G_a\,,
\ee
so that $d_V G[\xbar \xi^\alpha]$ reduces to the bulk term with undifferentiated $d_V \phi^\Gamma$ and $d_V \pi_\Gamma$,
\be
d_V G[\xbar \xi^\alpha] = \int d^d x \left((A_{\xbar \xi}^\Gamma + {A'}_{\xbar \xi}^\Gamma) d_V \pi_\Gamma -  d_V \phi^\Gamma (B_{\xbar \xi,\Gamma} + B'_{\xbar \xi,\Gamma} ) \right)\,,
\ee
as it should.  The on-shell value of $G[\xbar \xi^\alpha]$ is furthermore clearly equal to $F$. 

To illustrate the construction, consider  the case where $F$ is quadratic, 
\be
F = \frac12 g^{rs} Q_r Q_s + Q_r \oint d^{d-1}y g^{ra}(y) Q_a (y) + \frac12 \oint d^{d-1} y \oint d^{d-1}y' g^{ab}(y,y') Q_a(y) Q_b(y')\,,
\ee
where $g^{ab}(y,y')$  involves the delta function and its derivatives, e.g., $g^{ab}(y,y')= h^{ab}(y) \delta(y,y') + h^{abA}(y) \partial_A \delta(y,y')$, 
so that the double integral in $F$ reduces to the single integral 
$$\frac12 \oint d^{d-1} y Q_a \left(h^{ab} Q_b  + h^{abA} \partial_A Q_b \right).$$
The coefficients $g^{rs}$ and $g^{ab}(y,y')$ are symmetric by construction,
\begin{align}
g^{rs} &= g^{sr}\,, \\
g^{ab}(y,y') &= g^{ba}(y',y) \quad \Leftrightarrow \quad h^{ba} = h^{ab}-\partial_A{h^{abA}}, \; h^{abA} = - h^{baA}\,.
\end{align}
The asymptotic symmetry parameters are then
\begin{align}
\xbar U^s &= \frac {\partial F}{\partial Q_s} = g^{rs} Q_r + \oint d^{d-1}y g^{sa}(y) Q_a (y) \,,\\
\xbar T^a(y) &= \frac {\delta F}{\delta Q_a(y)} = g^{sa}(y) Q_s + h^{ab}(y) Q_b(y) + h^{abA} (y) \partial_A Q_b(y) \, .
\end{align}
Because $g^{ab}(y,y')$ is symmetric, if the $a$-component of the parameter of the asymptotic symmetry involves the $b$-component of the charge, the $b$-component of the parameter of the asymptotic symmetry will involve in turn the $a$-component of the charge.

We can draw two conclusions from the above analysis.
\begin{itemize}
\item There is no difficulty in handling non-linear expressions in the asymptotic symmetry charges, and these take the standard form of an asymptotic symmetry generator, i.e., are the sum of a standard bulk term proportional to the constraints and of the appropriate surface integral.  Furthermore, the asymptotic redefinition determines everything modulo physically irrelevant proper gauge symmetries.
\item Integrability of the charges is manifest in the analysis since we work with the integrated charges themselves throughout.  The corresponding transformation (Hamiltonian vector field) is derived by taking the Poisson bracket.  

Alternatively, one might ask the following question: given a transformation that preserves the boundary conditions on the fields, with gauge parameters $\xi^\alpha$ going asymptotically to  $$\overset{\circ}{\xi}\,^\alpha (r, y^A, \tilde U^s, \tilde T^a, \partial_A \tilde T ^a, \cdots)$$ where $\tilde U^s$ and $ \tilde T^a$ are a priori allowed to depend on the fields.  What are the restrictions on that dependence that must be imposed  in order to guarantee the existence of a generator? This is equivalent to requesting invariance of the action under the gauge  transformation which has an asymptotic behaviour dictated by $\tilde U^s$ and $ \tilde T^a$, i.e., to the demand  that that gauge transformation defines a true Hamiltonian vector field.  Repeating the above steps, one finds that the generator $G[\xbar \xi^\alpha] = \int d^d x \xbar \xi^\alpha \mathcal H_\alpha + \tilde F$, if it exists, must fulfill
\be
d_V \tilde F = \tilde U^s d_V Q_s + \oint d^{d-1}y \tilde T^a d_V \mathcal G_a\,,
\ee
i.e., $\tilde U^s d_V Q_s + \oint d^{d-1}y \tilde T^a d_V \mathcal G_a$ must be integrable.  Since this differential involves $d_V Q_s $ and  $d_V \mathcal G_a$ only, the same must be true for $\tilde U^s$, $ \tilde T^a$ and $\tilde F$.  Furthermore, integrability imposes that they obey the relations
\be
\tilde U^s = \frac {\partial \tilde F}{\partial Q_s}  \,, \qquad 
\tilde T^a(y) = \frac {\delta \tilde F}{\delta Q_a(y)}\, .
\ee
Under the condition of integrability, the functions of the charges exhaust all possibilities.
\end{itemize}

Finally, we note that in the case of gravity,  a definite choice of $\xi^\alpha$ corresponds to a definite choice of slicing of spacetime by spacelike hypersurfaces.  One sometimes speaks therefore of ``integrable slicings'' for the choice of $\xi^\alpha$ leading to well-defined charges \cite{Adami:2021nnf}.

\section{Redefinition of the BMS$_5$ algebra}
\label{sec:BMS5}

\subsection{Structure of the BMS$_5$ algebra}

In 5 spacetime dimensions, the generators of the BMS algebra include, besides the Poincar\'e generators, two types of supertranslations, namely,   the ``pure supertranslations'' and the ``subleading supertranslations''.  The pure supertranslations are diffeomorphisms of order $\mathcal O(1)$ as $r \rightarrow \infty$, with no $\ell \leq 1$ spherical harmonic components, which correspond to the standard translations.  They are the natural generalization to higher dimensions of the pure supertranslations in 4 spacetime dimensions.   The subleading supertranslations are diffeomorphisms of order $\mathcal O(r^{-1})$, which define improper gauge symmetries in $D=5$ dimensions thanks to the boundary conditions of \cite{Fuentealba:2021yvo,Fuentealba:2022yqt} which make them relevant.  They also do not involve $\ell \leq 1$ spherical harmonic components.   

{While at different orders in $r$ at $D=5$, the two supertranslation} branches are at the same order  $\mathcal O(1)$ at $D=4$.  Because of this coincidence, the second branch acquires  a logarithm.    In higher dimensions, the second branch decays as $\mathcal O(r^{-(D-4)})$ and is independent from the $\mathcal O(1)$ branch.

Denoting  the generators of the homogeneous Lorentz transformations, of the ordinary translations, of the pure supertranslations and of the subleading supertranslations respectively by $M_a$, $T_i$, $S_{\alpha}$ and $L^{\beta}$, one finds that the BMS$_5$ algebra has the structure  \cite{Fuentealba:2021yvo,Fuentealba:2022yqt} 
\begin{flalign}
\{M_a, M_b\} & = f_{ab}^c M_c\,, \label{eq:bracketMM5}\\
\{M_a, T_i\} & = R_{ai}^{\phantom{ai}j}T_j\,,\\
\{M_a, S_\alpha\} & = G_{a\alpha}^{\phantom{a\alpha}i}T_i + G_{a\alpha}^{\phantom{a\alpha}\beta}S_\beta + U_{a \alpha \beta \gamma}L^\beta L^\gamma\,, \quad U_{a \alpha \beta \gamma} = U_{a \alpha (\beta \gamma)} \,,\label{eq:bracketMS5}\\
\{M_a, L^\alpha\} & = - G_{a\beta}^{\phantom{a\alpha}\alpha}L^\beta\,, \label{eq:bracketML5}\\
\{L^\alpha, S_\beta\} & = \delta^\alpha_\beta\,. \label{eq:bracketLS5}
\end{flalign}
It is a non-linear algebra, with  quadratic terms appearing in the bracket $\{M_a, S_\alpha\}$.  The relation (\ref{eq:bracketLS5}) expresses that the generators of the subleading supertranslations $L^\alpha$ and the generators of the pure supertranslations $S_\beta$ are canonically conjugate.  It is this key feature that enables one to simplify the algebra.  The subleading supertranslations play thus in $D=5$  the role played in $D=4$ by the logarithmic supertranslations. 

The explicit expressions of the charges $M_a$, $T_i$, $S_{\alpha}$ and $L^{\beta}$ generating the asymptotic symmetries are given in \cite{Fuentealba:2021yvo,Fuentealba:2022yqt}.  Since the asymptotic symmetries are diffeomorphisms, each charge-generator takes the form
\be \label{eq:Charge}
\int d^4x \left(\xi \mathcal H + \xi^k \mathcal H_k \right) + \oint_{S^3_\infty} d^3x \mathcal B\,,
\ee
where $\mathcal H \approx 0$ and $\mathcal H_k \approx 0$ are the gravitational constraints and where $\oint_{S^3_\infty} d^3x \mathcal B$ is a surface integral on the three-sphere $S^3_\infty$ at infinity, involving the coefficients of the metric, of its conjugate momentum and of the vector field $(\xi,\xi^k)$ in the expansion near infinity.   The vector field $(\xi,\xi^k)$ {tends to a homogeneous Lorentz transformation for $M_a$, to an ordinary spacetime translation for $T_i$, to a pure supertranslation for $S_\alpha$ and to a logarithmic supertranslation for $L^\beta$.}

{As recalled above, the canonical generator \eqref{eq:Charge}} is the sum of (i) a bulk contribution proportional to the constraints multiplied by the components of the vector field parametrizing the corresponding diffeomorphism and (ii) a surface term involving the asymptotic form of this diffeomorphism, such that the key Hamiltonian relation $\iota_X \Omega = - d_V F_X$ relating charges $F_X$ and phase space Hamiltonian vector fields $X$ through the symplectic form $\Omega$, does indeed hold.  These expressions will not be needed here since our derivations rely only on the form (\ref{eq:bracketMM5})-(\ref{eq:bracketLS5}) of the algebra of the asymptotic symmetry generators, without involving the detailed form of the charges.

Jacobi identities constrain the structure constants. Among these, the ones that give useful relations for the subsequent construction involve the Lorentz generators and the pure supertranslations. From
\begin{equation}
\{M_{a},\{M_{b},S_{\alpha}\}\}+\{M_{b},\{S_{\alpha},M_{a}\}\}+\{S_{\alpha},\{M_{a},M_{b}\}\}=0\,,
\end{equation}
we get
\begin{align}
G_{b\alpha}^{\,\,\,\,\,j}R_{aj}^{\,\,\,\,\,i}+G_{b\alpha}^{\,\,\,\,\,\beta}G_{a\beta}^{\,\,\,\,\,i}-\left(a\leftrightarrow b\right) & =f_{ab}^{c}G_{c\alpha}^{\,\,\,\,\,i}\,,\\
G_{b\alpha}^{\,\,\,\,\,\beta}G_{a\beta}^{\,\,\,\,\,\gamma}-\left(a\leftrightarrow b\right) & =f_{ab}^{c}G_{c\alpha}^{\,\,\,\,\,\gamma}\,,\\
G_{b\alpha}^{\,\,\,\,\,\beta}U_{a\beta\gamma\delta}-2G_{a\delta}^{\,\,\,\,\,\beta}U_{b\alpha\gamma\beta}-\left(a\leftrightarrow b\right) & =f_{ab}^{c}U_{c\alpha\gamma\delta}\,.\label{eq:Jacobi3}
\end{align}
Similarly,  the identity
\begin{equation}
\{M_{a},\{S_{\alpha},S_{\beta}\}\}+\{S_{\alpha},\{S_{\beta},M_{a}\}\}+\{S_{\beta},\{M_{a},S_{\alpha}\}\}=0\,,
\end{equation}
 implies that
\begin{equation}
U_{a[\alpha\beta]\gamma}=0 \qquad \Leftrightarrow \qquad U_{a\alpha\beta\gamma} = U_{a(\alpha\beta)\gamma} \, .
\end{equation}
Consequently, since $U_{a\alpha[\gamma\delta]}=0$, the structure constants $U_{a\alpha\gamma\delta}$ are totally symmetric in their Greek indices, 
\be U_{a\alpha\gamma\delta} = U_{a(\alpha\beta\gamma)}=0\,. \ee 

\subsection{Redefinitions of the homogeneous Lorentz generators}

In our paper \cite{Fuentealba:2022yqt}, we tried to simplify the nonlinear BMS$_5$ algebra by considering a special type of redefinitions where only the supertranslations $S_\alpha$ were modified by quadratic terms.  We proved the existence of  obstructions within this restricted context. 

It turns that these obstructions are absent, and that the nonlinear  BMS$_5$ algebra can be linearized and transformed into a direct sum algebra, if one allows redefinitions of the Lorentz generators which involve both quadratic and cubic terms in the supertranslations and subleading supertranslation generators - a possibility not explored in  \cite{Fuentealba:2022yqt}. 

The searched-for redefinitions are simply
\begin{align}
\tilde{M}_{a} & =M_{a}-G_{a\beta}^{\,\,\,\,\,i}L^{\beta}T_{i}-G_{a\beta}^{\,\,\,\,\,\gamma}L^{\beta}S_{\gamma}-\frac{1}{3}U_{a\beta\gamma\delta}L^{\beta}L^{\gamma}L^{\delta}\\
 & =M_{a}-L^{\beta}\{M_{a},S_{\beta}\}+\frac{2}{3}U_{a\beta\gamma\delta}L^{\beta}L^{\gamma}L^{\delta}\, , \label{Eq:NewM2}
\end{align}
(other generators unchanged). {Note that the coefficients $U_{a\alpha\gamma\delta}$ are totally symmetrized in their Greek indices when multiplied by  $L^{\beta}L^{\gamma}L^{\delta}$ in (\ref{Eq:NewM2}), but since they are totally symmetric, no information on the $U$'s is lost in the expression, i.e., one can completely read off $U_{a\alpha\gamma\delta}$ from (\ref{Eq:NewM2}).}

One can easily prove that the brackets of these new Lorentz generators
with ordinary translations remain the same and that the brackets with
all supertranslations vanish:
\begin{align}
\{\tilde{M}_{a},T_{i}\} & =R_{ai}^{\,\,\,\,\,j}T_{j}\,,\\
\{\tilde{M}_{a},S_{\alpha}\} & =L^{\beta}\{\{M_{a},S_{\beta}\},S_{\alpha}\}+2U_{a\beta\gamma\delta}\{L^{\beta},S_{\alpha}\}L^{\gamma}L^{\delta}\nonumber \\
 & =-2U_{a\beta\gamma\delta}L^{\beta}\{L^{\gamma},S_{\alpha}\}L^{\delta}+2U_{a\beta\gamma\delta}\{L^{\beta},S_{\alpha}\}L^{\gamma}L^{\delta}\nonumber \\
 & =0\,,\\
\{\tilde{M}_{a},L^{\alpha}\} & =\{M_{a},L^{\alpha}\}-G_{a\beta}^{\,\,\,\,\,\gamma}L^{\beta}\{S_{\gamma},L^{\alpha}\}\nonumber \\
 & =\{M_{a},L^{\alpha}\}+G_{a\beta}^{\,\,\,\,\,\alpha}L^{\beta}\nonumber \\
 & =0\,.
\end{align}
Using  the consequences of the Jacobi identities, one can also prove that the redefinition of the Lorentz generators defines a nonlinear automorphism
of the Lorentz algebra, i.e., that the $\tilde{M}_{a}$'s obey the same algebra as the $M_a$'s, 
\begin{align}
\{\tilde{M}_{a},\tilde{M}_{b}\} & =\{M_{a}-G_{a\beta}^{\,\,\,\,\,i}L^{\beta}T_{i}-G_{a\beta}^{\,\,\,\,\,\gamma}L^{\beta}S_{\gamma},M_{b}-G_{b\mu}^{\,\,\,\,\,j}L^{\mu}T_{j}-G_{b\mu}^{\,\,\,\,\,\nu}L^{\mu}S_{\nu}\}\\
 & =\{M_{a},M_{b}\}+\left(G_{b\delta}^{\,\,\,\,\,\beta}G_{a\beta}^{\,\,\,\,\,\gamma}-G_{a\delta}^{\,\,\,\,\,\beta}G_{b\beta}^{\,\,\,\,\,\gamma}\right)L^{\delta}S_{\gamma}\nonumber \\
 & \quad-\left[G_{b\gamma}^{\,\,\,\,\,i}R_{ai}^{\,\,\,\,\,j}+G_{b\gamma}^{\,\,\,\,\,\beta}G_{a\beta}^{\,\,\,\,\,j}-G_{a\gamma}^{\,\,\,\,\,i}R_{bi}^{\,\,\,\,\,j}-G_{a\gamma}^{\,\,\,\,\,\beta}G_{b\beta}^{\,\,\,\,\,j}\right]L^{\gamma}T_{j}\nonumber \\
 & \quad+\left(U_{b\gamma\mu\nu}G_{a\beta}^{\,\,\,\,\,\gamma}-U_{a\gamma\mu\nu}G_{b\beta}^{\,\,\,\,\,\gamma}\right)L^{\beta}L^{\mu}L^{\nu}\\
 & =f_{ab}^{c}\left(M_{c}-G_{c\beta}^{\,\,\,\,\,i}L^{\beta}T_{i}-G_{c\beta}^{\,\,\,\,\,\gamma}L^{\beta}S_{\gamma}-\frac{1}{3}U_{c\beta\gamma\delta}L^{\beta}L^{\gamma}L^{\delta}\right)\nonumber \\
 & =f_{ab}^{c}\tilde{M}_{c}\,.
\end{align}

The new form of the algebra is therefore
\begin{flalign}
\{\tilde{M}_{a},\tilde{M}_{b}\} & = f_{ab}^c \tilde{M}_{c}\,,\\
\{\tilde{M}_{a},T_{i}\} & =R_{ai}^{\,\,\,\,\,j}T_{j}\,,\\
\{\tilde{M}_{a},S_{\alpha}\} & = 0 \,,\label{eq:bracketMS5B}\\
\{\tilde{M}_{a},L^{\alpha}\} & = 0\,, \label{eq:bracketML5B}\\
\{L^\alpha, S_\beta\} & = \delta^\alpha_\beta\,, \label{eq:bracketLS5B}
\end{flalign}
which explicitly exhibits the direct sum structure ``Poincar\'e $\oplus$ Supertranslations'' (including the subleading supertranslations).  This is exactly the same structure as found in four spacetime dimensions \cite{Fuentealba:2022xsz}.

\subsection{Redefined transformations}
The redefined Lorentz generators lead to redefined Lorentz transformations. The variations of the fields under the new Lorentz transformations are simply obtained by taking their bracket with the new charges.  Since the redefinitions of the charges are nonlinear, the added terms to the original transformations will correspond to supertranslations and subleading supertranslations with coefficients that depend on the fields through the charges (see Section {\bf \ref{Sec:AsymptoticNonlinear}}).

This yields, for any phase space function $F$,
\begin{align}
\{F, \tilde{M}_{a}\} = &\{F,M_{a}\} - G_{a\beta}^{\,\,\,\,\,i}L^{\beta}\{F, T_{i}\} - G_{a\beta}^{\,\,\,\,\,\gamma}L^{\beta}\{F, S_{\gamma}\} \nonumber\\
&-\left(G_{a\beta}^{\,\,\,\,\,i}T_{i} -G_{a\beta}^{\,\,\,\,\,\gamma}S_{\gamma}- U_{a\beta\gamma\delta}L^{\gamma}L^{\delta}\right)\{F, L^{\beta}\} \label{Eq:CompDiff}
\end{align}
This expression explicitly shows that in addition to the original Lorentz variation $\{F,M_{a}\}$, the transformation of $F$ involves an ordinary translation with coefficient  $-G_{a\beta}^{\,\,\,\,\,i}L^{\beta}$, a pure supertranslation with coefficient $-G_{a\beta}^{\,\,\,\,\,\gamma}L^{\beta}$ and a subleading supertranslation with coefficient $-(G_{a\beta}^{\,\,\,\,\,i}T_{i} -G_{a\beta}^{\,\,\,\,\,\gamma}S_{\gamma}- U_{a\beta\gamma\delta}L^{\gamma}L^{\delta})$.

By substituting in (\ref{Eq:CompDiff}) the explicit form of the charges given in  \cite{Fuentealba:2022yqt}, one obtains the expression of the compensating field-dependent diffeomorphisms (in particular, their relevant asymptotic behaviour) that must be added to the Lorentz transformations to give to the algebra the direct sum structure.

\section{Redefinition of the BMS$_4$ superalgebra}
\label{Sec:SUSYV1}

\subsection{Algebra and Jacobi identities}
We now turn to supergravity.
We start with the version of the super-BMS algebra found
in \cite{Fuentealba:2021xhn} where the asymptotic symmetries are the following:
\begin{itemize}
\item Bosonic asymptotic symmetries (with respective charge-generators $M_a$, $T_i$ and $S_\alpha$):
\begin{itemize}
\item diffeomorphisms asymptotically behaving at infinity as homogeneous Lorentz transformations
\item ordinary translations and 
\item
pure supertranslations.  
\end{itemize}
They evidently form a subalgebra isomorphic to the BMS$_4$ algebra. 
\item Fermionic asymptotic symmetries:
The fermionic asymptotic symmetries are of three different types.  First, there are the local supersymmetry transformations with parameters asymptotically approaching a Killing spinor (i.e., becoming asymptotically constant in asymptotically cartesian frames ).  Their charge-generators are denoted $Q_I$.  Then there are the local supersymmetry transformations with parameters asymptotically approaching higher spherical harmonics modes. The corresponding supercharges are in infinite number and denoted $q_A$.  Finally, there are the conjugate transformations acting on surface variables, with charge-generators $s^A$ \cite{Fuentealba:2021xhn,Fuentealba:2020aax}.  
\end{itemize}

The explicit expressions are given in \cite{Fuentealba:2021xhn} and are not needed here, because our derivation relies solely on the structure of the charge algebra, which is, 
\begin{align}
[M_{a},M_{b}] & =f_{ab}^{c}M_{c}\,,\\{}
[M_{a},T_{i}] & =R_{ai}^{j}T_{j}\,,\\{}
[M_{a},S_{\alpha}] & =G_{a\alpha}^{i}T_{i}+G_{a\alpha}^{\beta}S_{\beta}\,,\\{}
[M_{a},Q_{I}] & =g_{aI}^{J}Q_{J}+V_{aIB}^{i}s^{B}T_{i}+V_{aIB}^{\alpha}s^{B}S_{\alpha}\,,\\{}
[M_{a},q_{A}] & =h_{aA}^{B}q_{B}+U_{aAB}^{i}s^{B}T_{i}+U_{aAB}^{\alpha}s^{B}S_{\alpha}\,,\\{}
[M_{a},s^{B}] & =-h_{aC}^{B}s^{C}\,,\\
\{s^{A},q_{B}\} & =\delta_{B}^{A}\,,\\
\{Q_{I},q_{A}\} & =d_{IA}^{i}T_{i}+d_{IA}^{\alpha}S_{\alpha}\,,\\
\{q_{A},q_{B}\} & =d_{AB}^{i}T_{i}+d_{AB}^{\alpha}S_{\alpha}\,,\\
\{Q_{I},Q_{J}\} & =d_{IJ}^{i}T_{i}\,.
\end{align}
The remaining brackets involving translations and supertranslations
vanish
\begin{align}
[T_{i},T_j] &= [T_{i},S_\alpha]  = [S_\alpha, S_\beta] = 0 \, , \\
[S_{\alpha},Q_{I}] &=[T_{i},Q_{I}]= 0 \, , \\
[S_{\alpha},q_{A}] &=[T_{i},q_{A}]= 0 \, , \\
[S_{\alpha},s^{B}] & =[T_{i},s^{B}]=0\,,
\end{align}
and we have also
\be
\{Q_{I},s^{A}\} = 0 = \{s^A,s^{B}\} \, .
\ee
The asymptotic symmetry superalgebra is nonlinear since it involves quadratic terms in the brackets of the Lorentz generators with the supercharges $Q_I, q_A$.

The following Jacobi identities are particularly useful:
\begin{itemize} 
\item From
\begin{equation}
[M_{a},[M_{b},S_{\alpha}]]+[M_{b},[S_{\alpha},M_{a}]]+[S_{\alpha},[M_{a},M_{b}]]=0\,,
\end{equation}
we derive the relations
\begin{align}
G_{b\alpha}^{j}R_{aj}^{i}+G_{b\alpha}^{\beta}G_{a\beta}^{i}-\left(a\leftrightarrow b\right) & =f_{ab}^{c}G_{c\alpha}^{i}\,,\label{eq:Id-fGi}\\
G_{b\alpha}^{\beta}G_{a\beta}^{\gamma}-\left(a\leftrightarrow b\right) & =f_{ab}^{c}G_{c\alpha}^{\gamma}\,.\label{eq:Id-fGgamma}
\end{align}
\item Analogously, from
\begin{equation}
[M_{a},[M_{b},T_{i}]]+[M_{b},[T_{i},M_{a}]]+[T_{i},[M_{a},M_{b}]]=0\,,
\end{equation}
follows the relation
\begin{equation}
R_{bj}^{k}R_{ak}^{i}-\left(a\leftrightarrow b\right)=f_{ab}^{c}R_{cj}^{i}\,.\label{eq:Id-fR-RR}
\end{equation}
\item The identity
\begin{equation}
[M_{a},\{Q_{I},q_{A}\}]-\{q_{A},[M_{a},Q_{I}]\}+\{Q_{I},[q_{A},M_{a}]\}=0\,,
\end{equation}
enables one to express the coefficients $V$'s of the quadratic terms in terms of the other structure constants,
\begin{align}
V_{aIB}^{i} & =d_{IB}^{\alpha}G_{a\alpha}^{i}+d_{IB}^{j}R_{aj}^{i}-g_{aI}^{J}d_{JB}^{i}-h_{aB}^{C}d_{IC}^{i}\,,\label{eq:Id-Vi}\\
V_{aIB}^{\beta} & =d_{IB}^{\alpha}G_{a\alpha}^{\beta}-g_{aI}^{J}d_{JB}^{\beta}-h_{aB}^{C}d_{IC}^{\beta}\,.\label{eq:Id-Vbeta}
\end{align}
\item Similarly, from
\begin{equation}
[M_{a},\{q_{A},q_{B}\}]-\{q_{B},[M_{a},q_{A}]\}+\{q_{A},[q_{B},M_{a}]\}=0\,,
\end{equation}
we can express the $U$'s,
\begin{align}
2U_{a(AB)}^{i} & =d_{AB}^{\,\,\,\,\,\alpha}G_{a\alpha}^{i}+d_{AB}^{j}R_{aj}^{i}-h_{aB}^{C}d_{AC}^{i}-h_{aA}^{C}d_{BC}^{i}\,,\label{eq:Id-Ui}\\
2U_{a(AB)}^{\beta} & =d_{AB}^{\alpha}G_{a\alpha}^{\beta}-h_{aB}^{C}d_{AC}^{\beta}-h_{aA}^{C}d_{BC}^{\beta}\,.\label{eq:Id-Ubeta}
\end{align}
\item Finally, the identity
\begin{equation}
[M_{a},[M_{b},q_{A}]]+[M_{b},[q_{A},M_{a}]]+[q_{A},[M_{a},M_{b}]]=0\,,
\end{equation}
implies the relations
\begin{align}
h_{bA}^{C}h_{aC}^{B}-\left(a\leftrightarrow b\right) & =f_{ab}^{c}h_{cA}^{B}\,,\label{eq:Id-fh-hh}\\
h_{bA}^{C}U_{aCB}^{i}-h_{aB}^{\,\,\,\,\,C}U_{bAC}^{i}+R_{aJ}^{i}U_{bAB}^{j}+G_{a\alpha}^{i}U_{bAB}^{\alpha}-\left(a\leftrightarrow b\right) & =f_{ab}^{c}U_{cAB}^{i}\,,\label{eq:Id-fUi}\\
h_{bA}^{\,\,\,\,\,C}U_{aCB}^{\alpha}-h_{aB}^{C}U_{bAC}^{\alpha}+G_{a\beta}^{\alpha}U_{bAB}^{\beta}-\left(a\leftrightarrow b\right) & =f_{ab}^{c}U_{cAB}^{\alpha}\,.\label{eq:Id-fUalpha}
\end{align}
\end{itemize}

\subsection{Algebraic decoupling of the fermionic charges}

In order to decouple the infinite-dimensional set of BMS conjugate supercharges
$q_{A}$ and $s_A$  in the algebra, we must redefine not only the Lorentz generators but also the supercharges $Q_I$ and $q_A$. The following nonlinear redefinitions achieve the requested task:
\begin{align}
\tilde{Q}_{I} & =Q_{I}-d_{IB}^{i}s^{B}T_{i}-d_{IB}^{\alpha}s^{B}S_{\alpha}\,, \label{Eq:RedSusy1}\\
\tilde{q}_{A} & =q_{A}-\frac{1}{2}d_{AB}^{i}s^{B}T_{i}-\frac{1}{2}d_{AB}^{\alpha}s^{B}S_{\alpha}\,, \label{Eq:RedSusy2}\\
\tilde{M}_{a} & =M_{a}+h_{aA}^{B}s^{A}q_{B}+\frac{1}{2}\left(U_{aAB}^{i}-h_{aA}^{C}d_{BC}^{i}\right)s^{A}s^{B}T_{i}+\frac{1}{2}\left(U_{aAB}^{\alpha}-h_{aA}^{C}d_{BC}^{\alpha}\right)s^{A}s^{B}S_{\alpha}\,.\label{Eq:RedSusy3}
\end{align}

Using the above consequences of the Jacobi identities, it is indeed straightforward, although somewhat tedious, to check  that:
\begin{itemize}
\item The new supercharges $\tilde Q_I$ obey the same algebra as the $Q_I$'s,
\begin{equation}
\{\tilde{Q}_{I},\tilde{Q_{J}}\}=\{Q_{I},Q_{J}\}=d_{IJ}^{i}T_{i}\,.
\end{equation}
and furthermore have vanishing Poisson bracket with the $\tilde{q}_{A}$'s
\be
\{\tilde{Q}_{I},\tilde{q}_{A}\} =0 \, .
\ee
In addition one has
\be
\{\tilde{q}_{A},\tilde{q}_{B}\}  =0\,,
\ee
and of course, the brackets
\be
\{\tilde Q_{I},s^{A}\} = 0 = \{s^A,s^{B}\}, \qquad  \, , \{s^{A},\tilde q_{B}\}  =\delta_{B}^{A}
\ee
are unchanged.
\item The brackets between the fermionic and Lorentz generators get simplified to
\be
[\tilde{M}_{a},\tilde{Q}_{I}] = g_{aI}^{J}\tilde{Q}_{J} \, , \qquad [\tilde{M}_{a},\tilde{q}_{C}]= 0 = [\tilde{M}_{a},s^{C}]\, .
\ee
\item The new Lorentz generators obey the Lorentz algebra,
\begin{equation}
[\tilde{M}_{a},\tilde{M}_{b}]=f_{ab}^{c}\tilde{M}_{c}\,. \label{Eq:NewLorentzS0}
\end{equation}
\end{itemize}
The derivation of all these relations is direct and not particularly illuminating.  To give an idea of the arguments being used, we provide the explicit verification of (\ref{Eq:NewLorentzS0}) in the appendix.

We have thus proved that after the redefinitions (\ref{Eq:RedSusy1})-(\ref{Eq:RedSusy3}) are made, the super-BMS algebra reduces to
\begin{align}
[\tilde{M}_{a},\tilde{M}_{b}] & =f_{ab}^{c}\tilde{M}_{c}\,,\\{}
[\tilde{M}_{a},T_{i}] & =R_{ai}^{j}T_{j}\,,\\{}
[\tilde{M}_{a},\tilde{Q}_{I}] & =g_{aI}^{J}\tilde{Q}_{J}\,,\\{}
\{\tilde{Q}_{I},\tilde{Q}_{J}\} & =d_{IJ}^{i}T_{i}\,, \quad
[\tilde{Q}_{I},T_{i}]  =0\,, \quad [T_i,T_{j}]  =0 \, ,
\end{align}
(super-Poincar\'e algebra),
\begin{align}
[S_\alpha, S_\beta] & = 0 \, , \quad  [S_{\alpha},\tilde{q}_{A}] = 0 = [S_{\alpha},s^{B}]\\
\{s^{A},\tilde{q}_{B}\} & =\delta_{B}^{A}\,, \\
\{\tilde{q}_{A},\tilde{q}_{B}\} & =0 = \{s^A,s^{B}\}\,,
\end{align}
(infinite-dimensinal abelian superalgebra with a central extension in the fermionic sector).  The brackets between the two algebra all vanish, except the ones between the Lorentz generators and the pure supertranslations, which are unchanged,
\be
[\tilde{M}_{a},S_{\alpha}]  =G_{a\alpha}^{i}T_{i}+G_{a\alpha}^{\beta}S_{\beta}\,. \ee
We have thus not reached the direct sum structure yet. We will now show how to achieve this task.

\section{The log-BMS$_4$ superalgebra}
\label{Sec:SUSYV2}

\subsection{Log-BMS$_4$ superalgebra and Jacobi identities}

In the preceding developments, we did not include logarithmic supertranslations.  This is because the boundary conditions of \cite{Fuentealba:2021xhn} froze those transformations, which did not appear therefore as asymptotic symmetries.

Due to the absence of logarithmic supertranslations in the super-BMS$_4$ algebra of \cite{Fuentealba:2021xhn}, one can only disentangle the higher order supersymmetric transformations from the algebra, but not the ordinary supertranslations which do not have their canonically conjugate partners.

In order to complete the direct sum decomposition of the superalgebra, we need to bring in the logarithmic supertranslations.  This is achieved by relaxing the boundary conditions of \cite{Fuentealba:2021xhn} by allowing terms that are generated by diffeomorphisms involving asymptotically $\log r$, as in \cite{Fuentealba:2022xsz}.

We will not go through the tedious procedure of constructing these boundary conditions and the asymptotic symmetries and charges, but will give instead directly the resulting (nonlinear) algebra,   which is the minimal extension
that respects Jacobi identities.  It reads 
\begin{align}
[M_{a},M_{b}] & =f_{ab}^{c}M_{c}\,,\\{}
[M_{a},T_{i}] & =R_{ai}^{j}T_{j}\,,\\{}
[M_{a},S_{\alpha}] & =G_{a\alpha}^{i}T_{i}+G_{a\alpha}^{\beta}S_{\beta}\,,\\{}
[M_{a},L^{\alpha}] & =-G_{a\beta}^{\alpha}L^{\beta}\,,\\{}
[L^{\alpha},S_{\beta}] & =\delta_{\beta}^{\alpha}\,,\\{}
[M_{a},Q_{I}] & =g_{aI}^{J}Q_{J}+V_{aIB}^{i}s^{B}T_{i}+V_{aIB}^{\alpha}s^{B}S_{\alpha}\,,\\{}
[M_{a},q_{A}] & =h_{aA}^{B}q_{B}+U_{aAB}^{i}s^{B}T_{i}+U_{aAB}^{\alpha}s^{B}S_{\alpha}\,,\\{}
[M_{a},s^{B}] & =-h_{aC}^{B}s^{C}\,,\\
\{s^{A},q_{B}\} & =\delta_{B}^{A}\,,\\
\{Q_{I},q_{A}\} & =d_{IA}^{i}T_{i}+d_{IA}^{\alpha}S_{\alpha}\,,\\
\{q_{A},q_{B}\} & =d_{AB}^{i}T_{i}+d_{AB}^{\alpha}S_{\alpha}\,,\\
\{Q_{I},Q_{J}\} & =d_{IJ}^{i}T_{i}\,.
\end{align}
Jacobi identities $\left(L^{\alpha},q_{A},q_{B}\right)$ and $\left(L^{\alpha},q_{A},Q_{I}\right)$
require the presence of the additional non-vanishing brackets
\begin{align}
[L^{\alpha},q_{A}] & =n_{AB}^{\alpha}s^{B}\,,\quad[L^{\alpha},Q_{I}]=n_{IB}^{\alpha}s^{B}\,,
\end{align}
with
\begin{align}
d_{IA}^{\alpha} & =n_{IA}^{\alpha}\,,\qquad d_{AB}^{\alpha}=n_{AB}^{\alpha}+n_{BA}^{\alpha}\,.\label{eq:Id-d-n}
\end{align}
The remaining brackets vanish
\begin{equation}
[T_{i},Q_{I}]=[T_{i},q_{A}]=[S_{\alpha},q_{A}]=[L^{\alpha},Q_{I}]=[T_{i},s^{B}]=[S_{\alpha},s^{B}]=\{Q_{I},s^{A}\}=0\,.
\end{equation}

The following Jacobi identities are particularly useful:
\begin{itemize}
\item From
\begin{equation}
[M_{a},[M_{b},S_{\alpha}]]+[M_{b},[S_{\alpha},M_{a}]]+[S_{\alpha},[M_{a},M_{b}]]=0\,,
\end{equation}
we obtain the relations
\begin{align}
G_{b\alpha}^{j}R_{aj}^{i}+G_{b\alpha}^{\beta}G_{a\beta}^{i}-\left(a\leftrightarrow b\right) & =f_{ab}^{c}G_{c\alpha}^{i}\,,\label{eq:Id-fGi2}\\
G_{b\alpha}^{\beta}G_{a\beta}^{\gamma}-\left(a\leftrightarrow b\right) & =f_{ab}^{c}G_{c\alpha}^{\gamma}\,,\label{eq:Id-fGgamma2}
\end{align}
\item Analogously, from
\begin{equation}
[M_{a},[M_{b},T_{i}]]+[M_{b},[T_{i},M_{a}]]+[T_{i},[M_{a},M_{b}]]=0\,,
\end{equation}
we get the following relation
\begin{equation}
R_{bj}^{k}R_{ak}^{i}-\left(a\leftrightarrow b\right)=f_{ab}^{c}R_{cj}^{i}\,,\label{eq:Id-fR-RR2}
\end{equation}
\item The identity
\begin{equation}
[M_{a},\{Q_{I},q_{A}\}]-\{q_{A},[M_{a},Q_{I}]\}+\{Q_{I},[q_{A},M_{a}]\}=0\,,
\end{equation}
enables one again to express the quadratic structure constants $V_{aIB}^{i}$ and $V_{aIB}^{\beta}$ in terms of the other structure constants,
\begin{align}
V_{aIB}^{i} & =d_{IB}^{\alpha}G_{a\alpha}^{i}+d_{IB}^{j}R_{aj}^{i}-g_{aI}^{J}d_{JB}^{i}-h_{aB}^{C}d_{IC}^{i}\,,\label{eq:Id-Vi2}\\
V_{aIB}^{\beta} & =d_{IB}^{\alpha}G_{a\alpha}^{\beta}-g_{aI}^{J}d_{JB}^{\beta}-h_{aB}^{C}d_{IC}^{\beta}\,,\label{eq:Id-Vbeta2}
\end{align}
\item And similarly, the identity
\begin{equation}
[M_{a},\{q_{A},q_{B}\}]-\{q_{B},[M_{a},q_{A}]\}+\{q_{A},[q_{B},M_{a}]\}=0\,,
\end{equation}
leads to
\begin{align}
2U_{a(AB)}^{i} & =d_{AB}^{\alpha}G_{a\alpha}^{i}+d_{AB}^{j}R_{aj}^{i}-h_{aB}^{C}d_{AC}^{i}-h_{aA}^{C}d_{BC}^{i}\,,\label{eq:Id-Ui2}\\
2U_{a(AB)}^{\beta} & =d_{AB}^{\alpha}G_{a\alpha}^{\beta}-h_{aB}^{C}d_{AC}^{\beta}-h_{aA}^{C}d_{BC}^{\beta}\,,\label{eq:Id-Ubeta2}
\end{align}
\item The identity
\begin{equation}
[M_{a},[M_{b},q_{A}]]+[M_{b},[q_{A},M_{a}]]+[q_{A},[M_{a},M_{b}]]=0\,,
\end{equation}
implies
\begin{align}
h_{bA}^{C}h_{aC}^{B}-\left(a\leftrightarrow b\right) & =f_{ab}^{c}h_{cA}^{B}\,,\label{eq:Id-fh-hh2}\\
h_{bA}^{C}U_{aCB}^{i}-h_{aB}^{C}U_{bAC}^{i}+R_{aj}^{i}U_{bAB}^{j}+G_{a\alpha}^{i}U_{bAB}^{\alpha}-\left(a\leftrightarrow b\right) & =f_{ab}^{c}U_{cAB}^{i}\,,\label{eq:Id-fUi2}\\
h_{bA}^{C}U_{aCB}^{\alpha}-h_{aB}^{C}U_{bAC}^{\alpha}+G_{a\beta}^{\alpha}U_{bAB}^{\beta}-\left(a\leftrightarrow b\right) & =f_{ab}^{c}U_{cAB}^{\alpha}\,.\label{eq:Id-fUalpha2}
\end{align}
\item Finally, from the identities  involving the sets of generators
$\left(M_{a},L^{\alpha},q_{A}\right)$ and $\left(M_{a},S_{\alpha},q_{A}\right)$,
we can express the structure constants $U_{aAB}^{\alpha} $ and $V_{aIB}^{\alpha}$ in terms of the others,
\begin{align}
U_{aAB}^{\alpha} & =n_{AB}^{\beta}G_{a\beta}^{\alpha}-h_{aA}^{C}n_{CB}^{\alpha}-h_{aB}^{C}n_{AC}^{\alpha}\,,\label{eq:Id-Ualpha2}\\
V_{aIB}^{\alpha} & =n_{IB}^{\beta}G_{a\beta}^{\alpha}-g_{aI}^{J}n_{JB}^{\alpha}-h_{aB}^{C}n_{IC}^{\alpha}\,.
\end{align}
\end{itemize}

\subsection{Algebraic decoupling}

The redefinitions that enable one to algebraically decouple the infinite-dimensional set of charges $\left(S_{\alpha},q_{A},L^{\beta},s^{B}\right)$
from the super-Poincar\'e generators in the log-BMS$_4$ superalgebra are direct generalizations of those of the previous sections.  
We define
\begin{align}
\tilde{Q}_{I} & =Q_{I}-d_{IB}^{i}s^{B}T_{i}-d_{IB}^{\alpha}s^{B}S_{\alpha}\,,\\
\tilde{q}_{A} & =q_{A}-\frac{1}{2}d_{AB}^{i}s^{B}T_{i}-\frac{1}{2}d_{AB}^{\alpha}s^{B}S_{\alpha}\,,\\
\tilde{M}_{a} & =M_{a}-G_{a\beta}^{i}L^{\beta}T_{i}-G_{a\beta}^{\gamma}L^{\beta}S_{\gamma}+h_{aA}^{B}s^{A}q_{B} \nonumber \\
&+\frac{1}{2}\left(U_{aAB}^{i}-h_{aA}^{C}d_{BC}^{i}-n_{AB}^{\beta}G_{a\beta}^{i}\right)s^{A}s^{B}T_{i}-h_{aA}^{C}n_{CB}^{\alpha}s^{A}s^{B}S_{\alpha}\,. \\
\tilde{L}^\alpha &= L^\alpha + \frac12 n_{[AB]}^{\alpha}s^A s^{B}\,.
\end{align}

Using the above consequences of the Jacobi identities, we find successively that:
\begin{itemize}
\item The brackets of the fermionic generators $\{\tilde{Q}_{I},\tilde{Q}_{J}\}$, $\{\tilde{Q}_{I},\tilde{q}_{A}\}$ and $\{\tilde{q}_{A},\tilde{q}_{B}\}$ simplify to
\begin{equation}
\{\tilde{Q}_{I},\tilde{Q}_{J}\}=d_{IJ}^{i}T_{i}\, , \qquad \{\tilde{Q}_{I},\tilde{q}_{A}\}= 0\, , \qquad \{\tilde{q}_{A},\tilde{q}_{B}\}= 0\,,
\end{equation}
while the brackets
\be
\{\tilde Q_{I},s^{A}\} = 0 = \{s^A,s^{B}\} \, , \qquad   \{s^{A},\tilde q_{B}\}  =\delta_{B}^{A}\,,
\ee
are unchanged.
\item The brackets between the fermionic and Lorentz
generators reduce to
\begin{align}
[\tilde{M}_{a},\tilde{Q}_{I}] & = g_{aI}^{J}\tilde{Q}_{J}\,, \\
[\tilde{M}_{a},\tilde{q}_{C}] & = 0\, , \quad [\tilde{M}_{a},s^{C}]  = 0\,.
\end{align}
\item The brackets of the new logarithmic supertranslation generators $\tilde L^{\beta}$
with the super-Poincaré generators $\tilde{M}_{a}$, $T_{i}$ and
$\tilde{Q}_{I}$ are zero:
\begin{align}
[\tilde{M}_{a},\tilde L^{\alpha}] & =0\,, \\
[\tilde L^{\alpha},\tilde{Q}_{I}] &  =0\,.
\end{align}
Similarly,
\be [\tilde L^{\alpha},\tilde{q}_{A}] =0 \, .
\ee
\item Finally one verifies that the Lorentz algebra is fulfilled,
\begin{equation}
[\tilde{M}_{a},\tilde{M}_{b}]=f_{ab}^{c}\tilde{M}_{c}\,. \label{Eq:NewLorentzS1}
\end{equation}
\end{itemize}
Again, the derivation of all these relations is direct but rather tedious. We provide the explicit verification of (\ref{Eq:NewLorentzS1}) in the appendix.

Therefore, after the redefinitions given in this
section are performed, the log-BMS$_4$ superalgebra takes the direct sum structure of ``super-Poincar\'e'' $\oplus$ ``Abelian superalgebra with central extension'' (Heisenberg algebra).  The latter is infinite-dimensional and contains the pure supertranslations and the angle-dependent supersymmetries, which are thus decoupled from the super-Poincar\'e generators in the superalgebra.

The 
non-vanishing brackets are
\begin{align}
[\tilde{M}_{a},\tilde{M}_{b}] & =f_{ab}^{c}\tilde{M}_{c}\,,\\{}
[\tilde{M}_{a},T_{i}] & =R_{ai}^{j}T_{j}\,,\\{}
[\tilde{M}_{a},\tilde{Q}_{I}] & =g_{aI}^{J}\tilde{Q}_{J}\,,\\{}
\{\tilde{Q}_{I},\tilde{Q}_{J}\} & =d_{IJ}^{i}T_{i}\,, 
\end{align}
(super-Poincar\'e algebra) and 
\begin{align}
[L^{\alpha},S_{\beta}] & =\delta_{\beta}^{\alpha}\,,\\{}
\{s^{A},\tilde{q}_{B}\} & =\delta_{B}^{A}\,,
\end{align}
(infinite-dimensional Heisenberg algebra).

\section{Conclusions}
\label{sec:Conclusions}

In this paper, we have shown that the method of \cite{Fuentealba:2022xsz} can be extended to the BMS algebra in 5 dimensions and to BMS superalgebras.  In all cases, we have been able to redefine (super-)Poincar\'e generators that are free from pure supertranslation ambiguities and angle-dependent supergauge ambiguities.  Namely, we have exhibited new (super-)Poincar\'e generators that have vanishing Poisson brackets with the charge-generators of the pure supertranslations and of the angle-dependent supersymmetry transformations.

What makes this construction possible is the key property that the BMS supertranslations and BMS supersymmetry transformations have canonically conjugate charges, which play the same role as the logarithmic supertranslation charges of \cite{Fuentealba:2022xsz}.  Since the presence of canonical partners to the pure BMS transformations is expected to hold in higher dimensions and for extended supersymmetries (we verified this to be the case in the linear case), similar redefinitions are then also expected to be possible in those generalizations. 

{The redefinitions of the generators necessary for bringing the algebra to the desired direct sum form are nonlinear.  The nonlinear context is natural for discussing charges and symmetries and is compatible with the general structure $ \int d^d x \xi^\alpha \mathcal H_\alpha +B_\xi$ of generators of improper gauge transformations: the nonlinearities occur through the asymptotic nonlinear dependence of the gauge parameters on the charges.}

{We note that in the new presentations of the superalgebras, there is no square root of the pure supertranslations. We also note that since the BMS (super-)algebras can be written as direct sums of the (super-)Poincar\'e algebra with infinite-dimensional Heisenberg superalgebras, the theory of their representations takes a simpler form.  We hope to return to this question in the future.  }

\section*{Acknowledgments}
We thank C\'edric Troessaert for important discussions.  O.F. is grateful to the Collège de France for kind hospitality while this work was completed. This work was partially supported by a Marina Solvay Fellowship (O.F.) and by  FNRS-Belgium (conventions FRFC PDRT.1025.14 and IISN 4.4503.15), as well as by funds from the Solvay Family.

\appendix

\section{Brackets of new Lorentz generators}
\label{AppendixA}

In this appendix, we provide the detailed derivation of the relation (\ref{Eq:NewLorentzS1}).  If one sets the logarithmic supertranslation generators equal to zero, this yields also a derivation of (\ref{Eq:NewLorentzS0}).

Our aim is to prove the relation
\begin{equation}
[\tilde{M}_{a},\tilde{M}_{b}]=f_{ab}^{c}\tilde{M}_{c}\,.
\end{equation}
The explicit computation goes as follows. After manipulation of the
indices and using the algebra, we obtain that
\begin{align}
[\tilde{M}_{a},\tilde{M}_{b}] & =f_{ab}^{c}\tilde{M}_{c}+\left[-f_{ab}^{d}h_{dB}^{C}-h_{bD}^{C}h_{aB}^{D}+h_{aD}^{C}h_{bB}^{D}\right]s^{B}q_{C}\label{eq:MM1}\\
 & \quad-\left[\frac{1}{2}f_{ab}^{d}\left(U_{dBC}^{i}-h_{dB}^{A}d_{CA}^{i}-n_{BC}^{\beta}G_{d\beta}^{i}\right)+h_{aB}^{D}h_{bC}^{E}d_{DE}^{i}\right]s^{B}s^{C}T_{i}\nonumber \\
 & \quad+\Big[h_{bB}^{D}U_{aDC}^{i}+\frac{1}{2}\left(U_{bBC}^{\alpha}-h_{bB}^{D}d_{CD}^{\alpha}-G_{b\beta}^{\alpha}n_{BC}^{\beta}\right)G_{a\alpha}^{i} \nonumber \\
 & \quad +\frac{1}{2}\left(U_{bBC}^{j}-h_{bB}^{A}d_{CA}^{j}-n_{BC}^{\beta}G_{b\beta}^{j}\right)R_{aj}^{i}-\left(a\leftrightarrow b\right)\Big]s^{B}s^{C}T_{i}\label{eq:MM2}\\
 & \quad-\left[f_{ab}^{d}\left(-h_{dB}^{D}n_{DC}^{\alpha}\right)+h_{aB}^{D}h_{bC}^{E}d_{DE}^{\alpha}\right]s^{B}s^{C}S_{\alpha}\nonumber \\
 & \quad+\left[h_{bB}^{D}U_{aDC}^{\alpha}+\left(-h_{bB}^{D}n_{DC}^{\alpha}\right)G_{a\beta}^{\alpha}-\left(a\leftrightarrow b\right)\right]s^{B}s^{C}S_{\alpha}\,.\label{eq:MM3}
\end{align}
The second term at the right hand side of \eqref{eq:MM1} vanishes
by virtue of the identity \eqref{eq:Id-fh-hh2}. Using the identity
\eqref{eq:Id-Ui2}, all the terms in the line above \eqref{eq:MM2}
together with the terms in the line \eqref{eq:MM2} reduce to
\begin{align*}
\frac{1}{2}\left(f_{ab}^{d}h_{dB}^{D}+h_{bF}^{D}h_{aB}^{F}-h_{aF}^{D}h_{bB}^{F}\right)d_{CD}^{i}s^{B}s^{C}T_{i}\\
-\frac{1}{2}f_{ab}^{d}U_{d[AB]}^{i}s^{A}s^{B}T_{i}+\left[2h_{bA}^{C}U_{a[CB]}^{i}+U_{b[AB]}^{j}R_{aj}^{i}+U_{b[AB]}^{\alpha}G_{a\alpha}^{i}-\left(a\leftrightarrow b\right)\right]s^{A}s^{B}T_{i}\\
\frac{1}{2}n_{AB}^{\beta}f_{ab}^{d}G_{d\beta}^{i}s^{A}s^{B}T_{i}+\frac{1}{2}n_{AB}^{\beta}\left[-G_{b\beta}^{\alpha}G_{a\alpha}^{i}-G_{b\beta}^{j}R_{aj}^{i}-\left(a\leftrightarrow b\right)\right]s^{A}s^{B}T_{i}\,.
\end{align*}
The first line of the above expression is zero by virtue of the identity
\eqref{eq:Id-fh-hh2}. The second one vanishes by using the identity
\eqref{eq:Id-fUi2}, while the third one is zero after considering
indentity \eqref{eq:Id-fGi2}.

For the remaining terms in the line above \eqref{eq:MM3} and those
in \eqref{eq:MM3}, we make use of the identity \eqref{eq:Id-Ualpha2}.
These terms reduce then to
\begin{align*}
 & \left(f_{ab}^{d}h_{dB}^{D}+h_{bF}^{D}h_{aB}^{F}-h_{aF}^{D}h_{bB}^{F}\right)n_{DC}^{\alpha}s^{B}s^{C}S_{\alpha}\,,
\end{align*}
which vanishes by virtue of the identity \eqref{eq:Id-fh-hh2}.


\begin{thebibliography}{99}

%\cite{Fuentealba:2022xsz}
\bibitem{Fuentealba:2022xsz}
O.~Fuentealba, M.~Henneaux and C.~Troessaert,
``Logarithmic supertranslations and supertranslation-invariant Lorentz charges,''
JHEP \textbf{02}, 248 (2023)
doi:10.1007/JHEP02(2023)248
[arXiv:2211.10941 [hep-th]].

  %\cite{Bondi:1962px}
\bibitem{Bondi:1962px}
H.~Bondi, M.~G.~J.~van der Burg and A.~W.~K.~Metzner,
``Gravitational waves in general relativity. 7. Waves from axisymmetric isolated systems,''
Proc. Roy. Soc. Lond. A \textbf{269} (1962), 21-52
doi:10.1098/rspa.1962.0161

%\cite{Sachs:1962wk}
\bibitem{Sachs:1962wk}
R.~K.~Sachs,
``Gravitational waves in general relativity. 8. Waves in asymptotically flat space-times,''
Proc. Roy. Soc. Lond. A \textbf{270} (1962), 103-126
doi:10.1098/rspa.1962.0206

 %\cite{Sachs:1962zza}
\bibitem{Sachs:1962zza}
  R.~Sachs,
  ``Asymptotic symmetries in gravitational theory,''
  Phys.\ Rev.\  {\bf 128} (1962) 2851.
  %%CITATION = doi:10.1103/PhysRev.128.2851;%%
   
%\cite{Fuentealba:2020ghw}
\bibitem{Fuentealba:2020ghw}
O.~Fuentealba, M.~Henneaux, S.~Majumdar, J.~Matulich and C.~Troessaert,
``Asymptotic structure of the Pauli-Fierz theory in four spacetime dimensions,''
Class. Quant. Grav. \textbf{37} (2020) no.23, 235011
doi:10.1088/1361-6382/abbe6e
[arXiv:2007.12721 [hep-th]].

%\cite{Benguria:1976in}
\bibitem{Benguria:1976in}
R.~Benguria, P.~Cordero and C.~Teitelboim,
``Aspects of the Hamiltonian Dynamics of Interacting Gravitational Gauge and Higgs Fields with Applications to Spherical Symmetry,''
Nucl. Phys. B \textbf{122} (1977), 61-99
doi:10.1016/0550-3213(77)90426-6

 %\cite{Coleman:1967ad}
\bibitem{Coleman:1967ad}
S.~R.~Coleman and J.~Mandula,
``All Possible Symmetries of the S Matrix,''
Phys. Rev. \textbf{159} (1967), 1251-1256
doi:10.1103/PhysRev.159.1251

%\cite{Fuentealba:2023syb}
\bibitem{Fuentealba:2023syb}
O.~Fuentealba, M.~Henneaux and C.~Troessaert,
``Asymptotic symmetry algebra of Einstein gravity and Lorentz generators,''
[arXiv:2305.05436 [hep-th]].

%\cite{Mirbabayi:2016axw}
\bibitem{Mirbabayi:2016axw}
M.~Mirbabayi and M.~Porrati,
``Dressed Hard States and Black Hole Soft Hair,''
Phys. Rev. Lett. \textbf{117}, no.21, 211301 (2016)
doi:10.1103/PhysRevLett.117.211301
[arXiv:1607.03120 [hep-th]].

  %\cite{Bousso:2017dny}
\bibitem{Bousso:2017dny}
R.~Bousso and M.~Porrati,
``Soft Hair as a Soft Wig,''
Class. Quant. Grav. \textbf{34} (2017) no.20, 204001
doi:10.1088/1361-6382/aa8be2
[arXiv:1706.00436 [hep-th]].  

%\cite{Javadinezhad:2018urv}
\bibitem{Javadinezhad:2018urv}
R.~Javadinezhad, U.~Kol and M.~Porrati,
``Comments on Lorentz Transformations, Dressed Asymptotic States and Hawking Radiation,''
JHEP \textbf{01} (2019), 089
doi:10.1007/JHEP01(2019)089
[arXiv:1808.02987 [hep-th]].

%\cite{Javadinezhad:2022hhl}
\bibitem{Javadinezhad:2022hhl}
R.~Javadinezhad, U.~Kol and M.~Porrati,
``Supertranslation-invariant dressed Lorentz charges,''
JHEP \textbf{04} (2022), 069
doi:10.1007/JHEP04(2022)069
[arXiv:2202.03442 [hep-th]].

%\cite{Chen:2021szm}
\bibitem{Chen:2021szm}
P.~N.~Chen, M.~T.~Wang, Y.~K.~Wang and S.~T.~Yau,
``Supertranslation invariance of angular momentum,''
Adv. Theor. Math. Phys. \textbf{25} (2021) no.3, 777-789
doi:10.4310/ATMP.2021.v25.n3.a4
[arXiv:2102.03235 [gr-qc]].

%\cite{Chen:2021zmu}
\bibitem{Chen:2021zmu}
P.~N.~Chen, J.~Keller, M.~T.~Wang, Y.~K.~Wang and S.~T.~Yau,
``Evolution of Angular Momentum and Center of Mass at Null Infinity,''
Commun. Math. Phys. \textbf{386} (2021) no.1, 551-588
doi:10.1007/s00220-021-04053-7
[arXiv:2102.03221 [gr-qc]].

%\cite{Compere:2021inq}
\bibitem{Compere:2021inq}
G.~Comp\`ere and D.~A.~Nichols,
``Classical and Quantized General-Relativistic Angular Momentum,''
[arXiv:2103.17103 [gr-qc]].
  
%\cite{Compere:2023qoa}
\bibitem{Compere:2023qoa}
G.~Comp\`ere, S.~E.~Gralla and H.~Wei,
``An asymptotic framework for gravitational scattering,''
[arXiv:2303.17124 [gr-qc]].

  %\cite{Strominger:2013jfa}
\bibitem{Strominger:2013jfa}
  A.~Strominger,
  ``On BMS Invariance of Gravitational Scattering,''
  JHEP {\bf 1407} (2014) 152
  [arXiv:1312.2229 [hep-th]].
  %%CITATION = doi:10.1007/JHEP07(2014)152;%%
  
  %\cite{Strominger:2017zoo}
\bibitem{Strominger:2017zoo}
A.~Strominger,
``Lectures on the Infrared Structure of Gravity and Gauge Theory,''
[arXiv:1703.05448 [hep-th]].

%\cite{Donnay:2018neh}
\bibitem{Donnay:2018neh}
L.~Donnay, A.~Puhm and A.~Strominger,
``Conformally Soft Photons and Gravitons,''
JHEP \textbf{01} (2019), 184
doi:10.1007/JHEP01(2019)184
[arXiv:1810.05219 [hep-th]].
%


%\cite{Fuentealba:2021yvo}
\bibitem{Fuentealba:2021yvo}
O.~Fuentealba, M.~Henneaux, J.~Matulich and C.~Troessaert,
``Bondi-Metzner-Sachs Group in Five Spacetime Dimensions,''
Phys. Rev. Lett. \textbf{128} (2022) no.5, 051103
doi:10.1103/PhysRevLett.128.051103
[arXiv:2111.09664 [hep-th]].

%\cite{Fuentealba:2022yqt}
\bibitem{Fuentealba:2022yqt}
O.~Fuentealba, M.~Henneaux, J.~Matulich and C.~Troessaert,
``Asymptotic structure of the gravitational field in five spacetime dimensions: Hamiltonian analysis,''
JHEP \textbf{07} (2022), 149
doi:10.1007/JHEP07(2022)149
[arXiv:2206.04972 [hep-th]].

%\cite{Fuentealba:2021xhn}
\bibitem{Fuentealba:2021xhn}
O.~Fuentealba, M.~Henneaux, S.~Majumdar, J.~Matulich and T.~Neogi,
``Local supersymmetry and the square roots of Bondi-Metzner-Sachs supertranslations,''
Phys. Rev. D \textbf{104} (2021) no.12, L121702
doi:10.1103/PhysRevD.104.L121702
[arXiv:2108.07825 [hep-th]].

%\cite{Henneaux:2018gfi}
\bibitem{Henneaux:2018gfi}
M.~Henneaux and C.~Troessaert,
``Asymptotic symmetries of electromagnetism at spatial infinity,''
JHEP \textbf{05} (2018), 137
doi:10.1007/JHEP05(2018)137
[arXiv:1803.10194 [hep-th]].

%\cite{Henneaux:1992ig}
\bibitem{Henneaux:1992ig}
M.~Henneaux and C.~Teitelboim,
``Quantization of gauge systems,''
Princeton University Press (Princeton: 1992)

%\cite{deBoer:1995cqx}
\bibitem{deBoer:1995cqx}
J.~de Boer, F.~Harmsze and T.~Tjin,
``Nonlinear finite W symmetries and applications in elementary systems,''
Phys. Rept. \textbf{272} (1996), 139-214
doi:10.1016/0370-1573(95)00075-5
[arXiv:hep-th/9503161 [hep-th]].

%\cite{Henneaux:1999ib}
\bibitem{Henneaux:1999ib}
M.~Henneaux, L.~Maoz and A.~Schwimmer,
``Asymptotic dynamics and asymptotic symmetries of three-dimensional extended AdS supergravity,''
Annals Phys. \textbf{282} (2000), 31-66
doi:10.1006/aphy.2000.5994
[arXiv:hep-th/9910013 [hep-th]].

%\cite{Henneaux:2010xg}
\bibitem{Henneaux:2010xg}
M.~Henneaux and S.~J.~Rey,
``Nonlinear $W_{infinity}$ as Asymptotic Symmetry of Three-Dimensional Higher Spin Anti-de Sitter Gravity,''
JHEP \textbf{12} (2010), 007
doi:10.1007/JHEP12(2010)007
[arXiv:1008.4579 [hep-th]].

%\cite{Campoleoni:2010zq}
\bibitem{Campoleoni:2010zq}
A.~Campoleoni, S.~Fredenhagen, S.~Pfenninger and S.~Theisen,
``Asymptotic symmetries of three-dimensional gravity coupled to higher-spin fields,''
JHEP \textbf{11} (2010), 007
doi:10.1007/JHEP11(2010)007
[arXiv:1008.4744 [hep-th]].




%\cite{Afshar:2013vka}
\bibitem{Afshar:2013vka}
H.~Afshar, A.~Bagchi, R.~Fareghbal, D.~Grumiller and J.~Rosseel,
``Spin-3 Gravity in Three-Dimensional Flat Space,''
Phys. Rev. Lett. \textbf{111}, no.12, 121603 (2013)
doi:10.1103/PhysRevLett.111.121603
[arXiv:1307.4768 [hep-th]].

%\cite{Gonzalez:2013oaa}
\bibitem{Gonzalez:2013oaa}
H.~A.~Gonzalez, J.~Matulich, M.~Pino and R.~Troncoso,
``Asymptotically flat spacetimes in three-dimensional higher spin gravity,''
JHEP \textbf{09}, 016 (2013)
doi:10.1007/JHEP09(2013)016
[arXiv:1307.5651 [hep-th]].


%\cite{Henneaux:2015ywa}
\bibitem{Henneaux:2015ywa}
M.~Henneaux, A.~Perez, D.~Tempo and R.~Troncoso,
``Hypersymmetry bounds and three-dimensional higher-spin black holes,''
JHEP \textbf{08}, 021 (2015)
doi:10.1007/JHEP08(2015)021
[arXiv:1506.01847 [hep-th]].


%\cite{Fuentealba:2015wza}
\bibitem{Fuentealba:2015wza}
O.~Fuentealba, J.~Matulich and R.~Troncoso,
``Asymptotically flat structure of hypergravity in three spacetime dimensions,''
JHEP \textbf{10}, 009 (2015)
doi:10.1007/JHEP10(2015)009
[arXiv:1508.04663 [hep-th]].

%\cite{Henneaux:2015tar}
\bibitem{Henneaux:2015tar}
M.~Henneaux, A.~P\'erez, D.~Tempo and R.~Troncoso,
``Extended anti-de Sitter Hypergravity in $2+1$ Dimensions and Hypersymmetry Bounds,''
doi:10.1142/9789813144101\_0009
[arXiv:1512.08603 [hep-th]].

%\cite{Fuentealba:2020zkf}
\bibitem{Fuentealba:2020zkf}
O.~Fuentealba, H.~A.~Gonz\'alez, A.~P\'erez, D.~Tempo and R.~Troncoso,
``Superconformal Bondi-Metzner-Sachs Algebra in Three Dimensions,''
Phys. Rev. Lett. \textbf{126}, no.9, 091602 (2021)
doi:10.1103/PhysRevLett.126.091602
[arXiv:2011.08197 [hep-th]].







%\cite{Regge:1974zd}
\bibitem{Regge:1974zd}
T.~Regge and C.~Teitelboim,
``Role of Surface Integrals in the Hamiltonian Formulation of General Relativity,''
Annals Phys. \textbf{88} (1974), 286
doi:10.1016/0003-4916(74)90404-7


%\cite{Adami:2021nnf}
\bibitem{Adami:2021nnf}
H.~Adami, D.~Grumiller, M.~M.~Sheikh-Jabbari, V.~Taghiloo, H.~Yavartanoo and C.~Zwikel,
``Null boundary phase space: slicings, news \& memory,''
JHEP \textbf{11} (2021), 155
doi:10.1007/JHEP11(2021)155
[arXiv:2110.04218 [hep-th]].

%\cite{Fuentealba:2020aax}
\bibitem{Fuentealba:2020aax}
O.~Fuentealba, M.~Henneaux, S.~Majumdar, J.~Matulich and T.~Neogi,
``Asymptotic structure of the Rarita-Schwinger theory in four spacetime dimensions at spatial infinity,''
JHEP \textbf{02} (2021), 031
doi:10.1007/JHEP02(2021)031
[arXiv:2011.04669 [hep-th]].






 

\end{thebibliography}
\end{document}